\documentstyle[sprocl,epsf]{article}
\begin{document}
\hfill McGill/97-18 \\
\title{HELIOSEISMOLOGY, MSW AND THE SOLAR NEUTRINO PROBLEM}
\author{P. BAMERT}
\address{Physics Department, McGill University, 3600 University St.,\\
Montr\'eal, Qu\'ebec, Canada, H3A 2T8}
\maketitle\abstracts{
In this talk I summarize recent work done in collaboration with Cliff Burgess and
Denis Michaud \cite{b97}, in which we performed a detailed investigation of how 
solar neutrinos propagate through helioseismic waves.
We find that the MSW solar neutrino spectrum is not modified at all
in the presence of seismic waves.
This finding differs from earlier estimates mainly because most
helioseismic waves are too weak in the vicinity of the MSW resonance to be of relevance
for neutrino propagation. A special class of waves may however by subject to an instability
and potentially have very large amplitudes. These waves do have long wavelengths,
a situation for which the formalism employed in earlier analyses does not apply.
Our numerical simulation significantly reduces their influence on neutrino propagation.
}

\section{Introduction}

The MSW~\cite{msw} effective hamiltonian description of neutrino resonant conversion in matter 
provides, by altering the neutrino spectrum, an excellent solution to the solar neutrino problem
and incidentally the best fit to solar neutrino data~\cite{exp} to date. 
The promise that the solar neutrino spectrum will be more precisely determined in the future
spurred investigations of how the MSW mean-field description might by altered in the presence
of density fluctuations.

A very efficient mechanism for modifying the MSW spectrum has since been 
proposed~\cite{fluctI,fluctII},
It involves the diffuse damping of neutrinos from the coherent MSW evolution 
as they interact with density fluctuations, and it
could give rise to significant effects, provided perturbations with 
correlation length very short compared to the neutrino oscillation length and the 
density scale height are actually present in the vicinity of the MSW resonance point.

What has been missing so far is a plausible source for such delta-correlated 
fluctuations within the sun.
We argue that they cannot come about as a superposition of 
helioseismic waves, the only plausible source for density fluctuations deep in the solar
interior. Our investigation shows that only a few seismic modes
can possibly reach amplitudes large enough to be of relevance.
A random superposition of these modes, however, does give rise to 
fluctuations with long spatial correlation length --- a situation for which the formalism
used in~\cite{fluctI,fluctII} breaks down. 

Here we explicitly compute the wave profiles generated by these seismic modes and 
numerically evolve neutrinos through it, without making use of the short-correlation-length
approximation. We find that the standard MSW neutrino spectrum remains unaltered. 

Below we briefly summarize the approach used in earlier 
analyses~\cite{fluctI,fluctII}. Section 3 then explains why this approach 
is only valid in the short 
correlation length regime. Section 4 discusses density fluctuations in the sun. Our 
results and conclusions are summarized in section 5.

\section{The Story So Far}

Neutrino flavor evolution in a unperturbed matter background is described, in the 
interaction picture, by
\begin{equation}
{\partial\rho_f\over\partial t}=-i[V_{\rm VAC}+V_{\rm MSW}(t) ,\rho_f],
\label{eqmsw}
\end{equation}
where $\rho_f$ is the neutrino flavor density matrix, $V_{\rm VAC}\approx k
+{m^\dagger m\over 2k}+\dots $ is the 
vacuum mixing with $k$ and $m$ being the neutrino momentum and mass matrix
respectively and $V_{\rm MSW}\equiv\sqrt{2}G_Fg^en_e(t)$ 
denotes the effective interaction with the matter background
with $n_e(t)$ being the electron density, $g^e$=diag(1,0)
the charged current coupling matrix and $G_F$ Fermis constant. When integrated Eq.~(\ref{eqmsw}) gives rise to the 
familiar Parke formula~\cite{parke} for the MSW survival probability. 

In the presence of density fluctuations Eq.~(\ref{eqmsw}) is modified to become~\cite{fluctI}
\begin{eqnarray}
{\partial\rho_f\over\partial t}&=&-i[V_{\rm VAC}+\langle V_{\rm MSW}(t)\rangle ,\rho_f]
\label{eqmswfluc} \\
&&2G_F^2{\cal A}(t)\left[ (g^e)^2\rho_f+\rho_f(g^e)^2-2g^e\rho_fg^e\right]+{\cal O}(G_F^3),
\nonumber
\end{eqnarray}
where $\langle\dots\rangle$ denotes the average over an ensemble of density fluctuations, so that
$\langle n_e(t) \rangle$ is the mean electron density. The fluctuations are then contained
within $\delta n_e\equiv n_e-\langle n_e\rangle$. The correlation integral ${\cal A}(t)$ is given by
\begin{equation}
{\cal A}(t)\equiv\int_{t'}^t\langle\delta n_e(t)\delta n_e(\tau )\rangle d\tau ,
\label{eqcorr}
\end{equation}
where $t'$ denotes the initial time of neutrino evolution. When integrated Eq.~(\ref{eqmswfluc})
gives rise to the generalized Parke formula for the neutrino survival probability $P_e(t)$~\cite{fluctI}:
\begin{equation}
P_e(t)={1\over 2}+\left({1\over 2}-P_J\right)\lambda \cos{2\theta_m(t')} \cos{2\theta_m(t)}.
\label{eqgenparke}
\end{equation}
Here $\theta_m$ is the neutrino mixing angle in matter and
$P_J$ denotes the standard (nonadiabatic) MSW jump probability~\cite{parke}. The damping factor $\lambda$
is given by
\begin{equation}
\lambda\equiv\exp\left[ -2G_F^2\int_{t'}^t{\cal A}(t)\sin^22\theta_m(\tau )d\tau\right].
\label{eqlambda}
\end{equation}
Notice that in the absence of fluctuations, $\delta n_e=0$, ${\cal A}(t)$ vanishes and Eqs.~(\ref{eqgenparke})
and (\ref{eqmswfluc})
reduce to the standard Parke formula~\cite{parke}  and to Eq.~(\ref{eqmsw}) respectively.

\section{Dealing With Long Correlation Lengths}

The formalism described in the previous section is only valid as long as the correlation lengths
of the involved density fluctuations is much smaller than the neutrino oscillation length.
To see this it is useful to expand the electron density into a complete
set of, respectively uncorrelated, orthonormal functions:
\begin{equation}
n_e(t)\equiv\langle n_e(t)\rangle\left[ 1+\sum_n{\cal C}_n\phi_n(t)\right]
\label{eqbasis}
\end{equation}
where the coefficients ${\cal C}_n$ are assumed to be gaussian distributed with vanishing mean, $\langle
{\cal C}_n\rangle = 0$, and uncorrelated with each other, $\langle {\cal C}_n {\cal C}_m\rangle = {\cal D}_n
\delta_{nm}$. These basis functions could for example represent helioseismic waves.

For further illustration it is instructive to consider a one dimensional toy model, 
the 'cell' model, in which the density fluctuation modes $\phi_n(t)$
are cells of fixed length '$\ell$', aligned one after the other so that
$\langle\delta n_e(t)\delta n_e(\tau )
\rangle =\epsilon^2\langle n_e(t)\rangle\langle n_e(\tau )\rangle$
if $t$ and $\tau$ are both within the same cell
and zero otherwise. 

For this model the correlation coefficient approximates 
to ${\cal A}\sim \ell\epsilon^2n_e^2$ and is obviously growing with
$\ell$. Since the second order term of Eq.~(\ref{eqmswfluc}) 
grows with ${\cal A}$ the expansion on which this equation is
based will inevitably break down at one point. This becomes more 
transparent when one realizes that the involved  
expansion parameter is not the weak interaction, $G_F$, but rather 
a combination of weak interaction ($G_F$), 
amplitude of the perturbation ($\epsilon$) and correlation length~($\ell$). 

To get a feeling for when this happens we performed a numerical average over the
ensemble of density profiles. Since any given neutrino sees only {\it one} specific 
density profile as it passes through the sun,
we can compute its survival probability using the standard MSW evolution (without 
fluctuations), Eq.~(\ref{eqmsw}), taking $n_e=\langle n_e\rangle +\delta n_e$.
In doing so for a statistical set of density profiles $\delta n_e$ we get an ensemble
average for the neutrino survival probability which is valid even for long correlation 
length \footnote{To save computer time we used the standard Parke formula (without fluctuations)
instead of numerically integrating the full evolution equation Eq.~(\ref{eqmsw}). The difference
turns out to be negligible.}.

\begin{figure}
\begin{center}
\leavevmode
\hbox{%
\epsfxsize=4.0in
\epsfysize=2.6in
\epsfbox{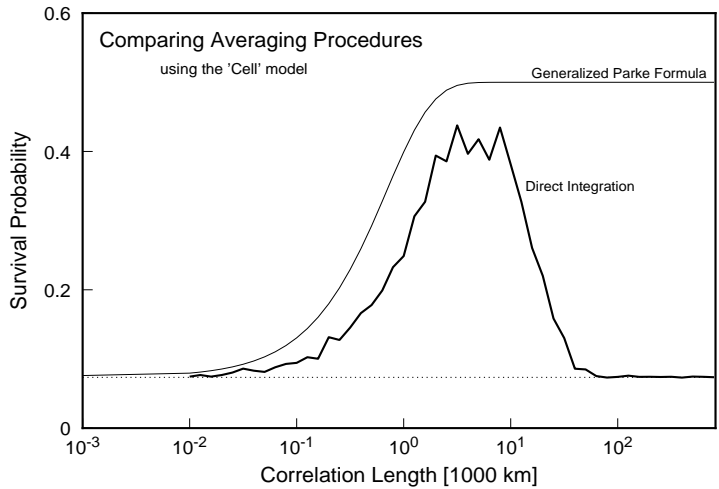}
}
\end{center}
\caption{Comparing the two averaging procedures in the case of the 'cell' model. The
neutrino parameters are: ${\delta m^2\over 2E}\sim 10^{-6} {\rm Ev}^2/{\rm MeV}$, $\sin^2
2\theta_V=0.01$ and $\epsilon^2=0.01$. The thin solid curve shows the result obtained using 
the generalized Parke formula Eq.(\ref{eqgenparke}), whereas the thick solid curves represents
the result of a direct numerical ensemble average. The horizontal line represents the standard MSW
survival probability, in the absence of fluctuations.} 
\label{figcomp}
\end{figure}

Where and how the pertrubative formalism, Eqs.~(\ref{eqmswfluc},\ref{eqgenparke}),
breaks down is illustrated in Fig.~(\ref{figcomp}), where we compared the two methods in the
framework of the 'cell' model described above.
The thin solid line shows the result obtained with the generalized Parke formula, 
Eq.~(\ref{eqgenparke}), whereas the thick solid line represents the method described above using a
statistical sample of 200 density profiles. 

As can be seen from this Figure the numerical
ensemble average begins to deviate significantly from the generalized Parke formula
for $\ell>$ a few $1000$ km. For even larger correlation length the 
only effect of the perturbation is a shift of the MSW resonance point, which doesn't affect 
the survival probability very much. Consequently the result of the numerical ensemble average 
approaches the standard MSW result for large correlation length.

From Fig.~(\ref{figcomp}) we learn two important points:

$\bullet$ The effect of fluctuations is largest when their correlation length is about
the length scale at which the perturbative approach breaks down, typically of the order of
the neutrino oscillation length.

$\bullet$ Using the generalized Parke formula beyond its domain of validity leads to a severe
overestimation of the influence of perturbations on neutrino propagation.

\section{Helioseismology And Density Fluctuations Inside The Sun}

Our question of concern is what kind of density perturbations in the sun (for a good review of
the solar interior see~\cite{solmod}) can affect
neutrino propagation. An inspection of~Eq.~(\ref{eqlambda}) shows that only density perturbations
in the vicinity of the MSW resonance point (where $\sin^22\theta_m$ is maximised) can significantly
modify the survival probability~\footnote{Even though Eq.~(\ref{eqgenparke}) only holds for
short correlation lengths this remark also proves true in the long correlation length regime.}.

Since the MSW resonance
typically occurs within the inner $0.5$ solar radii, density fluctuations in the outer shell
of the sun, {\it i.e.} the convective zone, and the surface won't be of relevance. Furthermore
a mixed, convective core has recently been excluded~\cite{mixedcore}. Barring the exotic option of
strange hydrodynamics, which is not described as perturbations on a spherically symmetric background
and which must not mix the core, deep inside the sun, this leaves us with 
helioseismic waves as a candidate for perturbations close to the MSW resonance point.

There are two basic types of helioseismic waves~\cite{solmod,solwaves}, 
called 'pressure' (p-) waves and 'buoyancy'
(g-) waves respectively, which owe their names to the main restoring force
that acts on a displaced element of matter, being pressure and gravity respectively.
Other features which distinguish the two types of waves are their periods (above/below 
30 minutes for g-/p- waves) and the regions which they populate.
Buoyancy waves typically are strongest in the central region of the sun whereas pressure
waves are most prominent in the convective zone.

Only pressure modes of low angular degree penetrate deeply into the solar interior. 
Helioseismic measurements however~\cite{soho} indicate that their amplitude is far to 
small to be of relevance to neutrino propagation~\cite{b97}. 

This leaves us with buoyancy waves, which typically have their largest amplitudes 
right in the vicinity of the MSW
resonance. Since they are damped in the gravitationally 
unstable convective zone they are generically hard to detect - and in fact haven't been
observed so far - and so could potentially be very strong, at least on purely phenomenological
grounds. However an inspection of the excitation and damping mechanisms indicates that, if at all,
only a few g-modes with radial wavenumber $n\leq 3$ can have large energies, at least 
$10^{35-37}$~Ergs~\cite{gmod}. This is because these waves could be linearly instable and so
would grow exponentially until nonlinearities saturate their growth~\cite{saturate}.
Such modes would correspond to $\delta n_e/n_e\sim 10^{-7}$, 
potentially giving rise to perturbations in the vicinity of $10^{-4,-5}$ when superimposed---
more if they were more energetic. 

\begin{figure}
\begin{center}
\leavevmode
\hbox{%
\epsfxsize=4.0in
\epsfysize=2.6in
\epsfbox{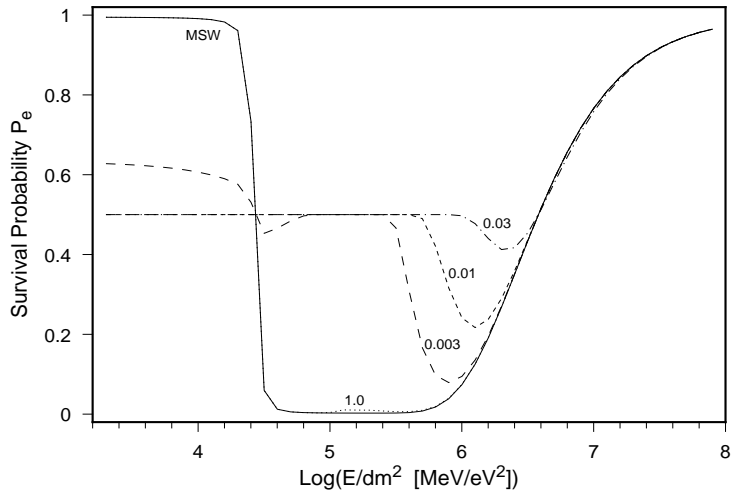}
}
\end{center}
\caption{Shown is the survival probability for electron neutrinos propagating through a
$\ell=13$, $n=1$ buoyancy wave. The various dashed and dot-dashed lines represent the results
obtained using the generalized Parke formula with wave amplitudes $\delta n_e/n_e$ as indicated.
Conversely the solid and dotted line show the result of a direct numerical integration, largely
indistinguishable from the standard MSW profile. For simplicity
of comparison all neutrinos have been assumed to be originating from the centre.} 
\label{figres}
\end{figure}

Since all of these instable modes have long correlation lengths and peak in the same area,
they will, when superimposed not give rise to delta correlated white noise
as has been assumed in earlier
works~\cite{fluctI,fluctII}. The formalism of section 2 does therefore not apply to them.

\section{Results And Conclusions}

We have computed the solar neutrino spectrum using density profiles generated from
these linearly instable g-modes. In doing so we averaged over neutrino production 
sites, using the production distributions of the Bahcall-Pinsonneault SSM~\cite{ssm},
and took into account details such as double resonant crossing.

We find that the solar neutrino spectrum is not affected on a measurable level by helioseismic
waves. This is shown in Fig.~(\ref{figres}) where we
have evolved the neutrinos through a helioseismic g-wave of angular degree $\ell=13$ and
radial order $n=1$. Fig.~(\ref{figres}) nicely illustrates the fact that applying the
generalized Parke formula can lead to a gross overestimate of the size of the effect.
The direct numerical integration shows no measurable effect at all even for a wave amplitude
as large as $100\%$.

The main reasons of why we find such a small effect are:

$\bullet$ Most seismic waves are much weaker than necessary to modify neutrino propagation.

$\bullet$ The few waves which could potentially be very strong have very long correlation lengths,
putting us in the regime where their influence on neutrino propagation is significantly reduced
(putting us in the right half of Fig.~(\ref{figcomp}))
compared to density fluctuations with shorter correlation lengths.

$\bullet$ Finally, averaging over production cites smears out part of the adiabatic dip of the 
MSW spectrum and provides an important background to any fluctuation effect.

We conclude that density perturbations in the sun are very unlikely to modify the
MSW neutrino spectrum and the MSW solution to the solar neutrino problem. 
They certainly
do not do so if they are composed of helioseismic waves.

\section*{Acknowledgements}
The author would like to thank Cliff Burgess and Denis Michaud
for the collaboration \cite{b97} on which this talk
is based. Furthermore we would like to acknowledge helpful conversations
with Joergen Christensen-Dalsgaard, Pawan Kumar and John Bahcall concerning
Helioseismology. This research was financially supported by NSERC of Canada.


\section*{References}
\begin{thebibliography}{999} 

\bibitem{b97}{P. Bamert, C.P. Burgess and D. Michaud, preprint 
McGill-97/13, {\tt hep-ph/9707542}, (1997).}

\bibitem{msw}{L. Wolfenstein, Phys. Rev. D {\bf 17}, 2369 (1978);\\
S.P. Mikheyev and A. Yu. Smirnov, {\it Sov. Phys. Usp.} {\bf 29} (1986) 1155;
Sov. J. Nucl. Phys. {\bf 42}, 913 (1985);
Nuovo Cimento  {\bf 9}, 17 (1986).}

\bibitem{exp}{R. Davis, D.S. Harmer and K.C. Hoffman, Phys. Rev. Lett. {\bf 20}, 1205 (1968);\\
J.K. Rowley {\it et al.}, in {\it Solar Neutrinos and Neutrino Astronomy},
AIP Conference Proceedings number 126, edited by M.L. Cherry, W.A. Fowler
and K. Lande, (1985); \\
K.S. Hirata {\it et al.}, Phys. Rev. Lett. {\bf 65}, 1297 (1990); \\
P. Anselmann {\it et al.}, Phys. Lett. B {\bf 327}, 377 (1994); \\
J.N. Abdurashitov {\it et al.}, Phys. Lett. B {\bf 328}{234} (1994).}

\bibitem{fluctI}{C. P. Burgess and D. Michaud, Ann. Phys. (N.Y.) {\bf 256}, 1 (1997);
and contribution to the proceedings of the 17th international
conference on Neutrino Physics and Astrophysics (NEUTRINO '96), ed by K. Enqvist, K. Huitu
and J. Maalampi, Helsinki, World Scientific, {\tt hep-ph/9611368}, (1996);\\
D. Michaud, McGill University M.Sc. thesis, 1994; in the 17th proceedings 
of the {\it Annual MRST Meeting}, (1995).}

\bibitem{fluctII}{F.N. Loreti and A.B. Balantekin, Phys. Rev. D {\bf 50}, 4762 (1994);\\
F.N. Loreti, Y.Z. Qian, G.M. Fuller and A.B. Balantekin, Phys. Rev. D {\bf 52}, 6664 (1995);\\
E. Torrente Lujan, preprint BUTP-96-8, {\tt hep-ph/9602398}, (1996);\\
H. Nunokawa, A. Rossi, V.B. Semikoz and J.W.F. Valle, 
Nucl. Phys. B {\bf 472}, 495 (1996);\\
A.B. Balantekin, J.M. Fetter and F.N. Loreti, Phys. Rev. D {\bf 54} 3941 (1996).}

\bibitem{parke}{S.J. Parke, Phys. Rev. Lett. {\bf 57}, 1275 (1986).}

\bibitem{solmod}{S. Turck-Chi\`eze {\it et al.}, Phys. Rep. {\bf 230}, 57 (1993).}

\bibitem{mixedcore}{J.N. Bahcall, M.H. Pinsonneault, S. Basu and J. Christensen-Dalsgaard, 
Phys. Rev. Lett {\bf 78}, 171 (1997).}

\bibitem{solwaves}{Jorgen Christensen-Dalsgaard, Lecture Notes,
available under http://bigcat.obs.aau.dk/$\sim$jcd/oscilnotes/.}

\bibitem{soho}{See e.g. B. Fleck, Rev. Mod. Astron. {\bf 10} (1997) in press, 
http://esa.nascom.nasa.gov /$\sim$bfleck/Preprints/ ;
see also ref~\cite{solmod} above.}

\bibitem{gmod}{P. Kumar, E.J. Quataert and J.N. Bahcall, Ap. J. {\bf 458}, L83 (1996).}

\bibitem{saturate}{P. Kumar and J. Goodman, Ap. J. in press, (1996).}

\bibitem{ssm}{J.N. Bahcall and M.H. Pinsonneault, Rev. Mod. Phys. {\bf 67}, 781 (1995).}

\end{thebibliography}
\end{document}